%% file: 00_main.tex
\documentclass[electronic]{vgtc}             

\graphicspath{{figures/}{pictures/}{images/}{./}} 

\usepackage{tabu}                      
\usepackage{booktabs}                  

\usepackage{lipsum}                    
\usepackage{mwe}                       

\usepackage{mathptmx}                  

\usepackage{multirow}

\usepackage{xcolor}
\usepackage{todonotes}
\presetkeys{todonotes}{color=blue!20}{}

\onlineid{0}

\vgtccategory{Research}

\vgtcinsertpkg

\title{
  Glow with the Flow: AI-Assisted Creation of Ambient Lightscapes for \\
  Music Videos
}

\author{
  Frederic Anthony Robinson\thanks{
    Authors contributed equally. Author order was decided by a random number generator.
  } 
  \and Vishnu Raj\footnotemark[1]
  \and David Cooper
  \and Fan Du
  \and David Gunawan
} %

\affiliation{
  Dolby Laboratories Inc.
  \thanks{
    \texttt{
      \{frederic.robinson,vishnu.raj,david.cooper,fan.du, david.gunawan\}
      @dolby.com
    }
  }
}

\teaser{
  \centering
  \includegraphics[width=0.95\linewidth]{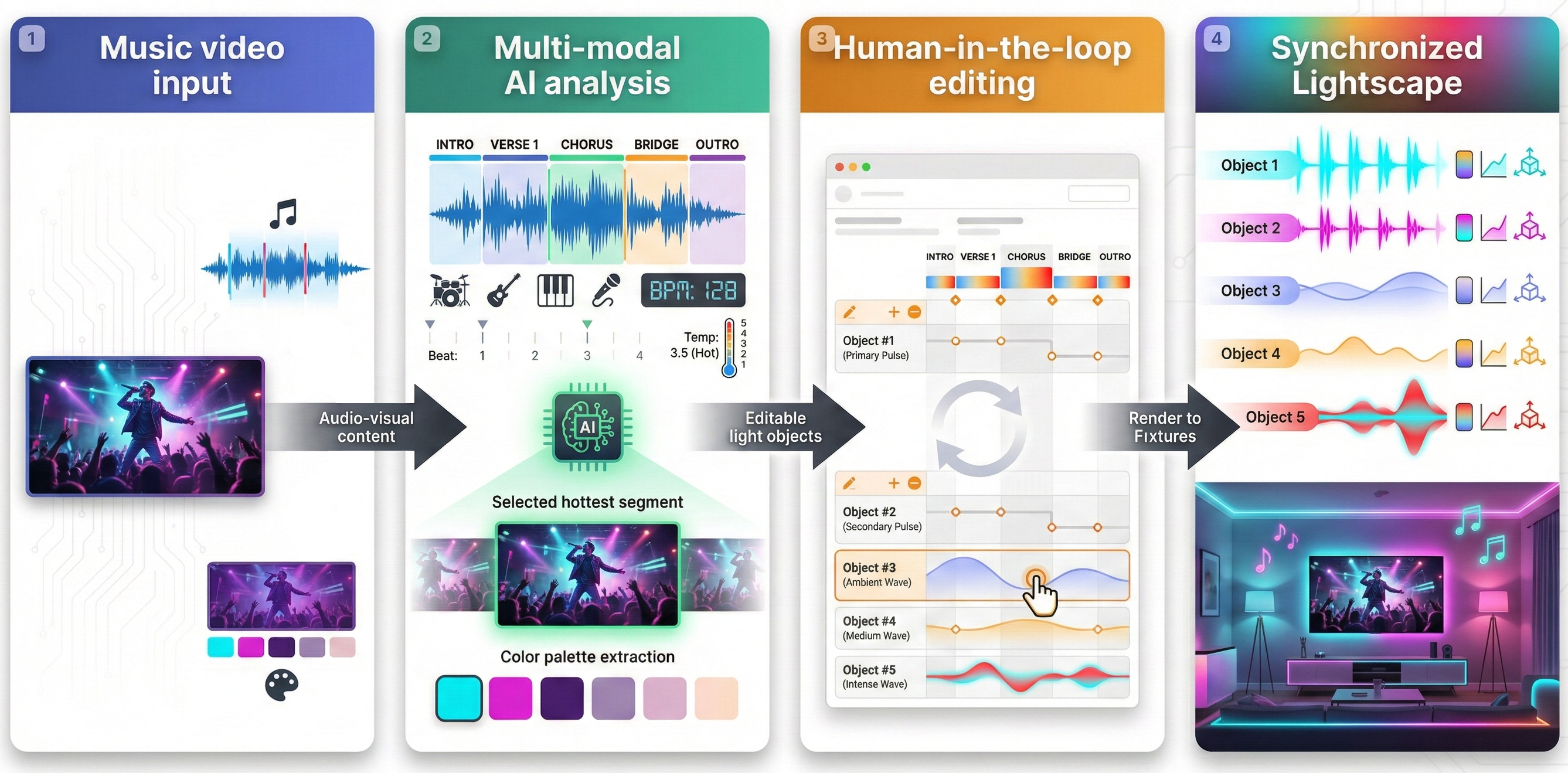}
  \caption{
    We present an AI-assisted method for creating ambient lightscapes for music videos using object based light representation. The system utilizes multi-modal AI to analyze music and visual content from a music video, extract key features and generates a synchronized lightscape in an editable object-based light show format. This enables human-in-loop refinement of light objects and effects through an intuitive authoring interface. The final output renders immersive, object-based lighting synchronized with the music video across various lighting fixtures.
  }
  \label{fig:teaser}
}

\abstract{
Designed light is an established modality for live performance and music playback. Despite the growing availability of consumer smart lighting, the creation of designed light for music visualization remains limited to professional contexts due to time and skill constraints. To address this, we present an AI-assisted system for generating ambient light sequences for music videos. Informed by professional design heuristics, the system extracts salient features from source video and audio to generate an editable preliminary design of object based ambient light effect. We evaluated the system by comparing its autonomous output against hand-authored designs for three music videos. Findings from responses by 32 participants indicate that the initial output provides a viable baseline for further refinement by human authors. This work demonstrates the utility of AI-assisted workflows in supporting the creation and adoption of designed light beyond professional venues.
    
} 


\begin{document}



\maketitle

\input{10_intro.tex}
\input{30_related.tex}

\input{50_method.tex}
\input{70_eval.tex}
\input{90_conclusion.tex}


\bibliographystyle{abbrv-doi}

\bibliography{99_library}
\end{document}

%% file: 10_intro.tex
\section{Introduction}

The visualization of music has traditionally been confined to the pixel grid. From early oscilloscopes to the media player visualizers of the late 1990s, the mapping was direct: audio amplitude often drove geometric distortion. In contrast, live music performances are accompanied by site-specific light designs created by experts for unique physical venues. The growing availability of affordable smart lighting for the home, and the gradual establishment of generalized IoT protocols promises to combine these fields, bringing immersive visualizations, comparable to venue experiences, to arbitrary environments, including the car and the home. One key barrier to this lies in the manual creation workflow, which creates a high barrier to entry regarding both expertise and time. Generating professional-grade light content requires specialized knowledge to conceptualize themes and translate audio-visual cues into compelling light events. Furthermore, the process is labor-intensive, necessitating manual analysis of timing cues and effect transitions. Consequently, the lack of trained practitioners and the inefficiency of current workflows hinder broader adoption. 

In response to these limitations, this work explores two fundamental inquiries: \emph{(a) What best practices do professional lighting designers employ to craft immersive lightscapes?}, and \emph{(b) To what extent can generative AI models codify and reproduce these strategies to support the authoring process of ambient light experiences?}

To address these questions, this paper introduces an AI-assisted system that supports the automated analysis of multimedia content and the synthesis of corresponding lightscape objects through multimodal processing. The proposed framework comprises two principal components:
\begin{itemize}
    \item \textbf{Salient cues identification}: A suite of AI algorithms parses audio and video streams to isolate perceptually significant attributes such as musical structure, rhythmic energy, and chromatic palettes, reflecting the attentional strategies commonly employed by professional lighting designers.
    \item \textbf{Generating light object mappings}: These extracted features are transformed into structured lighting cues through rule-based generative mappings grounded in expert design practices, enabling interpretable and controllable synthesis of ambient lightscapes.
\end{itemize}

The rest of this paper is organized as follows: in Sec. \ref{sec:related}, we discuss related work across music visualization, ambient light experiences and AI-assisted creation. Next, in Sec. \ref{sec:method}, we present our system and the professional practice its design is informed by. Finally, in Sec. \ref{sec:eval}, we discuss successes and limitations of our approach after evaluating the generated light designs by comparing them with hand-authored versions of three music videos.


%% file: 30_related.tex
\section{Related Works} \label{sec:related}
Research related to our work spans three closely connected domains: the visualization of music, the creation of ambient lighting, and AI-assisted creative workflows. Prior studies have examined how musical structure and affect can be mapped to visual representations, how ambient and architectural lighting can be designed to enhance media experiences, and how AI can support or automate creative decision-making.

\subsection{Visualization of music}
Traditionally, music visualizers have been largely \emph{semantically blind}, relying on low-level signal representations such as waveforms and frequency components without understanding instruments, structure, or emotional content. Early academic work focused on mapping audio features directly to visual forms in the time or frequency domain \cite{khulusi2020survey, lima2021survey}. While waveform plots, spectrograms, and time–frequency visualizations provide objective feedback for pitch, loudness, and timing, they fail to capture the semantic meaning of musical works \cite{djahwasi2024critical}. Efforts to move beyond these limitations include Hiraga et~al.’s exploration  \cite{hiraga2002performance} of visual interfaces that encode expressive performance features, mood, and musical structure through symbolic and qualitative representations. More recent systems, such as the audio-driven visualization framework by Graf et~al.~\cite{graf2021audio}, demonstrate that visuals derived from audio features can enhance perceived audiovisual complementarity compared to non-reactive visuals.

To move beyond passive waveform analysis, recent music visualization approaches increasingly employ machine learning to infer higher-level musical attributes and generate semantically aligned visuals. Prior work includes deep learning methods for sentiment-driven visual storytelling \cite{passalis2021deepsing}, multimodal systems that synchronize generated video content with musical style and rhythm using large language models \cite{huang2025deep}, and text-to-video frameworks such as Generative Disco \cite{liu2023generative} that produce beat-aligned visual narratives with interpretable design patterns . MuseDance \cite{dong2025every} further extends this paradigm with an end-to-end model that animates reference images from joint music and text inputs, enabling richer control over music-based video generation.


\subsection{Creating ambient light experiences}
Early work on ambient light experiences emphasized extending visual content into the viewer’s periphery to enhance immersion. IllumiRoom \cite{jones2013illumiroom} demonstrated effects such as \emph{field-of-view extension} and \emph{radial wobble} by augmenting a television with projection mapping, using a Kinect camera to project visuals onto surrounding walls and furniture. User studies showed that low-resolution peripheral cues, particularly motion and color, can significantly increase presence and immersion, despite limited visual acuity in peripheral vision. While IllumiRoom requires specialized hardware and careful calibration, similar atmospheric effects can be achieved more affordably through intelligent LED-based ambient lighting. Complementing this, Matviienko et~al.~\cite{matviienko2015towards} systematically characterized the ambient lighting design space across dimensions such as color, brightness, pattern, and spatial placement, providing a taxonomy for mapping light effects.

Subsequent work focused on how ambient light can convey information without distracting the user. Perteneder et~al.~\cite{perteneder2016glowworms} introduced the metaphors of \emph{Glowworms} for subtle, continuous lighting and \emph{Fireflies} for discrete, attention-grabbing signals, establishing HCI principles for attention management through controlled user studies. LightPlay \cite{fung2022lightplay} applied similar ideas in interactive media, replacing on-screen indicators in video games with ambient light strips and demonstrating faster user responses to peripheral light cues. More recently, research has shifted toward transferable, object-based lighting representations \cite{southwell2025creating}, in which abstract light objects are authored independently of specific fixtures and rendered at playback by a \emph{Flexible Light Renderer}. This approach decouples creative intent from hardware configuration, paralleling object-based audio systems such as Dolby Atmos, and enables ambient light experiences to be reproduced across diverse environments.
Recently, Skip-BART \cite{zhao2025automatic} is proposed as an end-to-end model that directly learns from experienced lighting engineers and predict vivid, human-like stage lighting but is limited to producing only a single color at any timepoints.

\subsection{AI Assisted creation}
Recent research increasingly frames AI as a collaborator within human-centered creative workflows rather than as a fully autonomous generator. Thoring et~al.~\cite{thoring2023augmented} articulate a framework for AI-supported design that emphasizes iterative exploration, explicit representation of design intent, and evolutionary creativity. In educational contexts, Schmitt-Fumian et~al.~\cite{schmitt2025ai} show that generative AI can function as a scalable feedback mechanism that enhances ideation and refinement while complementing human instruction. Tools such as FusAIn \cite{peng2025fusain} further operationalize this collaborative paradigm by enabling fine-grained, professional control over generative visual outputs through pen-based extraction and fusion of visual attributes. Extending AI-assisted creation to multimedia contexts, Mozualization \cite{xu2025mozualization} integrates text, images, and audio to support user-guided generation of music and visuals, emphasizing iterative adjustment and expressive alignment with user intent.  

Collectively, prior work reflects a clear shift toward AI as an active partner in creative processes, where effectiveness hinges on interaction design, representations, and feedback loops that preserve human agency. However, to the best of our knowledge, prior works such as Skip-BART \cite{zhao2025automatic}, are limited to generating single-color outputs and do not support iterative refinement or editable designs. In contrast, our work introduces the first AI-assisted workflow for authoring editable, object-based ambient lightscapes, positioning AI not as a replacement for human authorship but as a facilitator of co-evolutionary lighting design.

%% file: 50_method.tex
\section{AI-assisted lightscapes creation}  \label{sec:method}
\begin{figure*}
 \centering
 \includegraphics[width=\linewidth]{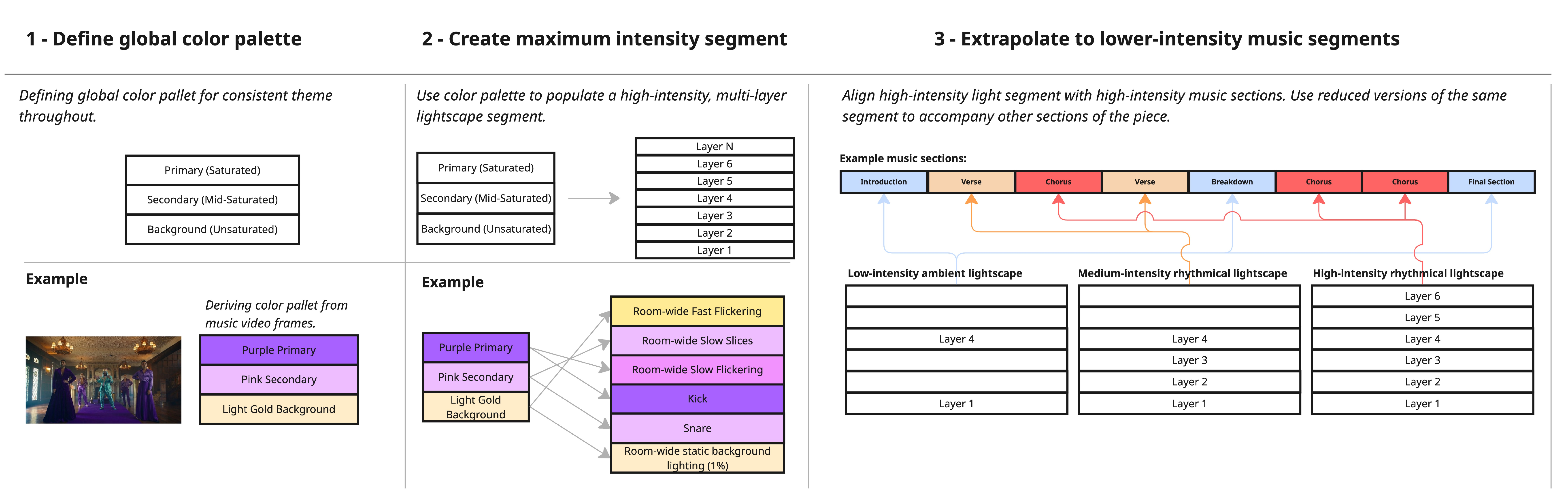}
 \caption{Mapping out the lightscape design process whereby a creative uses multimodal information from a music video to inform the creation of an accompanying light design. Video information determines an initial color palette which is then enriched to meet various functional requirements. Audio information on an event-, section-, and song-level then determines \textit{when} light events occur.}
 \label{fig:desig_process}
\end{figure*}

While an object-based format like that of \cite{southwell2025creating} solves the delivery problem, the creation of high-quality lightscapes remains a skill-intensive task. For a hand-crafted light show, a designer will first identify key elements of a given piece of music and then represent these using the light modality through programming cues for each available light fixture. This may, for example, mean accompanying snare drum hits with light flashes around the stage, or turning the entire venue deep red in sync with a guitar power chord.

\begin{figure}[!ht]
    \centering
    \includegraphics[width=\columnwidth]{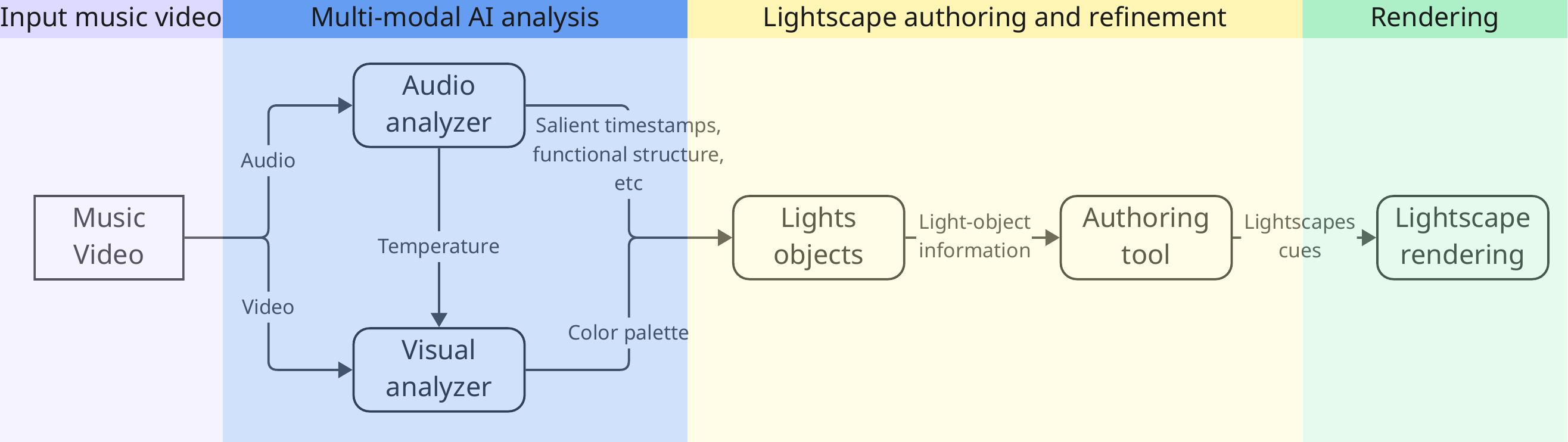}
    \caption{Conceptual flow of proposed system. On an input music video, we perform a multimodal AI which create cues for light objects. This light object representation in editable in authoring tool and rendered into the available light fixtures using \cite{southwell2025creating}.}
    \label{fig:proposed-system}
\end{figure}

When aiming to make this process \emph{AI-assisted}, the role of the algorithm shifts from a tool to a \emph{co-creative} partner. This necessitates a mechanism by which the semantic richness of a music video is similarly analyzed and down-sampled and then translated into light content that reflects its narrative or emotional essence. In this context, the AI must perform two complex tasks: understanding the semantic and structural content of the audio-video input (Analysis) and generating appropriate object-based lighting instructions (Synthesis) in an editable form. In this section, we explain in detail how the proposed system, shown in \cref{fig:proposed-system}, solves the challenges of analysis and synthesis, and in-turn presents a promising approach for assisted co-creation of ambient light experiences.

\input{53_method_modeling}

\input{55_method_analysis}
\input{57_method_synthesis}


%% file: 53_method_modeling.tex



\subsection{Modeling light designers' creative decision making}

We first consider how an expert light designer approaches a lightscape, based on informal conversations (formal analysis is reserved for future work). In essence, design decisions require a color palette to know \textit{what} colors to deploy, and audio events to determine \textit{when} to deploy them. A color palette may come from album cover art, music videos, or general palettes from the visual identity of a musical act. The music itself then determines the \textit{when}, both on a micro-level, i.e. individual audio events like a drum hit, and a macro-level, i.e. the overall progression of a piece of music throughout sections of a song. 

\cref{fig:desig_process} illustrates how a designer may extract relevant information from a music video to inform creative decisions on lighting. A global color palette is identified and used throughout the lightscape to maintain consistent coloring. The palette should cover both saturated and unsaturated colors, and may be based on a static image, such as an album cover, or from video frames of a music video. This color palette is then expanded to several related additional colors to design a high-intensity, multi-layer segment which then provides the blueprint for the highest-intensity section of the design. The segment is comprised of (i) sustained light layers whose parameters are modulated over time: FX Objects, and (ii) light events who are synchronized to individual musical elements: Audio Events. Finally, the segments of the music are analyzed and the previously compiled high-intensity segment is assigned to the high-intensity sections of the music. Reduced versions of the segment, meaning versions with fewer light elements and less activity, are then assigned to lower-intensity sections of the music. We refer to the level of intensity in a music segment as “temperature”.

The result of this process is a lightscape that (i) has a color palette that is aligned with that of the piece’s cover art or music video, and (ii) whose intensity adapts over time in line with the music’s progression. An automated system tasked with the same would thereby have to analyze audiovisual media to extract salient features and translate them into synchronized light events adjusted at event, section, and song levels.

%% file: 55_method_analysis.tex
\subsection{Multi-modal analysis of music videos} \label{subsec:analysis}


Analysis of a music video for salient information involves both audio analysis as well as visual analysis to derived corresponding light effects. Following the observations from \cite{graf2021audio} that audio-driven mappings were perceived to have higher audio-visual complementarity, we derive the timing ans intensity cues for the lighting effects primarily based on audio feature analysis, followed by color information from visual analysis. 

\noindent
\textbf{Audio analysis.} First, we perform audio analysis which involves a comprehensive music information retrieval to extract structural and rhythmic features from the audio content. The steps involved in audio processing are described below.
\begin{enumerate}
    \setlength{\itemsep}{1pt}
    \setlength{\parskip}{0pt}
    \setlength{\parsep}{0pt}
    \item \textbf{Stem separation.} As an initial step, the music track is decomposed into individual instruments to enable detailed analysis. Since access to original audio stems cannot be assumed in general music video workflows, we perform stem separation directly on the mixed audio using the state-of-the-art Demucs framework \cite{defossez2021hybrid,rouard2022hybrid}. Specifically, we employ the pretrained HT Demucs model \cite{rouard2022hybrid}to separate the mix into drums, vocals, bass, and remaining accompaniments.
    \item \textbf{Music information retrieval.} Music exhibits complex hierarchical structure, requiring multi-dimensional analysis for comprehensive understanding. For structural analysis, we adopt the method proposed in \cite{kim2023all}, which jointly performs beat and downbeat tracking along with functional structure segmentation and labeling. Using the separated audio sources, the model outputs three key descriptors: (i) BPM, representing the overall tempo; (ii) beat timestamps, providing precise temporal locations of beats; and (iii) functional segments, labeling high-level song sections such as intro, verse, chorus, bridge, solo, break, ending, and instrumental.
    \item \textbf{Dominant instrument event detection.} Peak events from dominant instruments act as primary rhythmic anchors for lighting, as beat accents and high-intensity transients are effective drivers of pulsing or flashing patterns that convey temporal dynamics. We therefore focus on drums, specifically kicks and snares, which define the rhythmic backbone of a track and provide clear temporal anchors for synchronization. LarsNet \cite{mezza2024toward} is used to separate kick and snare components from the drum stem, after which peak-picking algorithms extract precise timestamps for percussive events.
    \item \textbf{Temperature computation.} To modulate light intensity according to a track’s varying activity, we compute an energy-based \emph{temperature} score on a 1–5 scale for each functional segment. The score is derived from median RMS energy computed over short windows and normalized across the song, then quantized using K-means clustering with five centroids to ensure smooth, outlier-resistant assignments. These levels are categorized as \emph{cold} (low energy, 1–2), \emph{medium} (3–4), and \emph{hot} (high energy, 5).
\end{enumerate}

\noindent
\textbf{Visual analysis.} Visual analysis is critical for AI-driven lightscape creation, as it enables extraction of contextually appropriate color palettes that preserve aesthetic coherence with the video and ensure ambient lighting complements the visual narrative. Accordingly, our video analyzer derives color palettes from visually salient content through a multi-step process to inform ambient lighting design.

\begin{enumerate}
    \setlength{\itemsep}{1pt}
    \setlength{\parskip}{0pt}
    \setlength{\parsep}{0pt}
    \item \textbf{Salient Segment Identification.} The system identifies the most energetic segment by selecting the highest temperature value (level 5) from the audio analysis and designates it as the \emph{hottest segment}. This temperature-based method is one instantiation of a broader \emph{salient section identification} concept, which could alternatively be implemented using visual saliency detection, scene change analysis, or other content-driven cues.
    \item \textbf{Frame Processing \& Color Extraction.} Video frames from the identified segment are extracted and temporally subsampled to improve computational efficiency. A clustering-based color analysis is then applied to derive (i) a primary dominant color, (ii) a secondary prominent color, and (iii) a background color suitable for complementary or neutral ambient effects.
    \item \textbf{Soft Color Derivation.} The system generates ``soft" variants of primary and secondary colors by modifying saturation and value in HSV color space, creating five total color variations (primary, soft primary, secondary, soft secondary, background) for lighting design flexibility.
\end{enumerate}
This multi-modal approach prevents the generation of arbitrary or disconnected lighting schemes that would diminish immersion, instead producing ambient effects that feel organically integrated with both the visual and auditory elements of the multimedia content.

%% file: 57_method_synthesis.tex
\subsection{Synthesis of light objects}

The light object generator integrates audio and video analysis to produce synchronized lightscape objects across two temporal scales: event-level and section-level. Event-level lighting achieves rhythmic synchronization by responding to discrete audio events, primarily kick and snare hits, which serve as strong temporal anchors. Kick events trigger high-intensity flash effects using the primary video-derived color, while snare events use the secondary color, both precisely aligned to detected timestamps and shaped by configurable attack, hold, and release parameters to accentuate musical rhythm.

Section-level lighting provides evolving ambient context through layered effects driven by segment-wise temperature values. Low-energy segments (temperature 1–2) activate a subtle ambient layer with soft color variants and slow brightness modulation; medium-energy segments (3–4) introduce additional spatial modulation for moderate movement and intensity; and high-energy segments (5) engage all layers, adding dynamic pulsing effects with bright, desaturated background colors. A timeline is constructed from beat timestamps and functional segment boundaries, with keyframes generated accordingly and parameters interpolated for smooth transitions. The resulting light objects are encoded in an object-based lightscape format, enabling further human refinement or direct rendering to physical lighting systems \cite{southwell2025creating}.

%% file: 70_eval.tex
\section{Evaluation} \label{sec:eval}

To evaluate the system, we compare how well the system's output compares to hand-authored creations based on the same underlying media. While we envision our system to be used in co-creation contexts, we evaluate the system's unchanged ``first draft" to understand whether its output provides a helpful starting point for further manual refinement, or whether the system struggles to adequately implement professional designers' best practices. Does it save creators time so they can focus on the final details and polish?

\begin{figure}[tb]
 \centering
 \includegraphics[width=1\columnwidth]{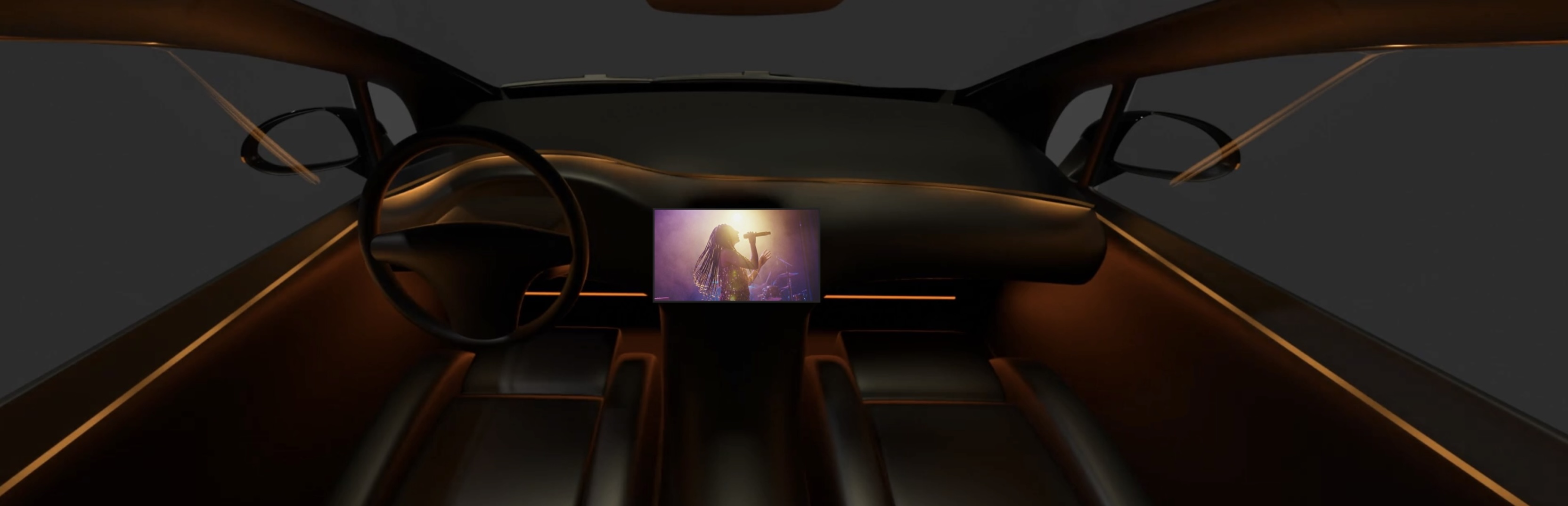}
 \caption{Light content visualization for a music video excerpt.}
 \label{fig:screenshot}
 \vspace*{-5mm}
\end{figure}

We had lighting artist create content for three music videos, and also auto-generated light content for the same using our system. We compiled five paired 10-second excerpts from each song, resulting in a total of 15 distinct musical segments in both a hand-authored and a generated version ($N=15$ pairs, referred as `contexts' from now on). These light designs were then visualized in an automotive environment (see Figure \ref{fig:screenshot}), producing a total of 30 video clips. All clips were processed identically aside from the lighting generation method. A side-by-side comparison of a simple example lightscape as hand-authored and AI-generation versions is in the supplementary materials.

\subsection{Study}
For an online survey, we recruited 32 participants to review the designs with no knowledge of which were hand-authored and which were generated. The participant pool was made up of 5 females and 27 males between the ages 25 and 50. 2 reported prior experience with lighting design and 3 with music production and 27 reported attending live music events either quarterly or biannually. 7 reported a Bachelor's Degree as highest education level, and 25 reported a Master's Degree or higher. After reading an introductory paragraph and providing demographic information, participants viewed paired stimuli (A and B) in randomized order. Raters then evaluated the clips immediately after viewing using a 10-point Likert scale. The evaluation criteria were 1) emotional congruence, 2) rhythmic accuracy, and 3) color match between video and lighting. The exact questions are shown in \cref{tab:questions}. They are informed by prior work by Zhao et. al \cite{zhao2025automatic}.

\begin{table}[h]
  \caption{List of Questions}
  \label{tab:questions}
  \centering
  \footnotesize
  \begin{tabular}{p{2.2cm} p{5.5cm}}
  \toprule
    Title & Question \\
  \midrule
    Emotional congruence (\texttt{emotional}) & How well does the light match the emotion of the music? \\
    Rhythmic synchronization (\texttt{rhythmic}) & How accurately does the lighting reflect the rhythm of the music? \\
    Chromatic lighting congruence (\texttt{chromatic}) & How well does the light match the color palette of the music video? \\
  \bottomrule
  \end{tabular}
  \vspace*{-5mm}
\end{table}

Ratings were collected across 15 contexts, with each participant evaluating both \emph{AI first pass} and \emph{human-authored} outputs. To avoid pseudo-replication, ratings were first averaged across participants for each context. For each perceptual attribute (emotional congruence, rhythmic synchronization, and chromatic congruence), AI and human version ratings were compared using two-sided Wilcoxon signed-rank tests. Holm–Bonferroni correction was applied to account for multiple comparisons. Effect sizes were quantified using rank-biserial correlation.

\subsection{Results and Discussion}
\begin{table}[!h]
  \caption{
    Results of Wilcoxon signed-rank tests comparing AI first pass and human-authored versions of lightscapes across perceptual attributes. $p_{holm}$ corresponds to the  Holm–Bonferroni correction. No attribute shows a statistically significant difference after correction for multiple comparisons.
  }
  \label{tab:wilcoxon_stats}
  \scriptsize%
  \centering%
  \begin{tabu}{r * {4}{c}}
    \toprule
    Attribute & $W$ & $p$ & $p_{holm}$ & $r$ \\
    \midrule
    \texttt{emotional} & 40.5 & 0.2769 & 0.5537 & 0.6625 \\
    \texttt{rhythmic}  & 49.5 & 0.5995 & 0.5995 & 0.5875 \\
    \texttt{chromatic} & 19.0 & 0.0181 & 0.0542 & 0.8417  \\
    \bottomrule
  \end{tabu}
\end{table}
\vspace*{-3mm}

\begin{table}[!h]
  \caption{Summary of statistics for $N = 15$ after averaging over 32 participants. Aggregated ratings were closely aligned across all three attributes, with differences remaining small (within approximately 0.1–0.3 points on a 10-point scale), indicating comparable central tendencies between the two conditions.
}
  \label{tab:ratings_stats}
  \scriptsize%
  \centering%
  \begin{tabu}{r * {6}{c}}
    \toprule
    \multirow{2}{*}{Version} 
      & \multicolumn{2}{c}{\texttt{emotional}}
      & \multicolumn{2}{c}{\texttt{rhythmic}}
      & \multicolumn{2}{c}{\texttt{chromatic}} \\
      & \rotatebox{90}{Mean} 
      & \rotatebox{90}{Median} 
      & \rotatebox{90}{Mean} 
      & \rotatebox{90}{Median} 
      & \rotatebox{90}{Mean} 
      & \rotatebox{90}{Median} \\
    \midrule
    Human         & 7.11 & 7.19 & 7.20 & 7.25 & 7.00 & 7.03 \\
    AI first pass & 7.00 & 7.06 & 7.27 & 7.31 & 7.27 & 7.28 \\
    \bottomrule
  \end{tabu}
\end{table}
\vspace*{-3mm}

\cref{tab:wilcoxon_stats} summarizes the results of Wilcoxon signed-rank tests comparing AI first pass and human-authored outputs across perceptual attributes. For each attribute (emotional congruence, rhythmic synchronization, and chromatic congruence), the table reports the Wilcoxon statistic (W), uncorrected p-value, Holm-corrected p-value and rank-biserial effect size (r). Ratings were averaged across participants for each context prior to analysis, and statistical significance was assessed using Holm-corrected p-values. 
\cref{tab:ratings_stats} shows that mean and median ratings for human-authored content and AI-generated version.

These tests revealed no statistically significant differences between human-authored lightscapes and AI first pass for any of the evaluated attributes after Holm correction. For emotional congruence, human-authored content is rated slightly higher than AI-generated lightscape (AI: Median = $7.06$; Human: Median = $7.19$), but the difference was not significant ($W = 40.5$, $p_{{holm}} = 0.55$, $r = 0.66$). Similarly, rhythmic synchronization showed no significant difference, with marginally higher ratings for AI-generated versions (AI: Median = $7.31$; Human: Median = $7.25$; $W = 49.5$, $p_{{holm}}$ = $0.60$, $r = 0.59$).

Chromatic congruence exhibited a strong trend favoring AI-generated version (AI: Median = $7.28$; Human: Median = $7.03$), reaching nominal significance prior to correction ($p = 0.018$), but not after Holm adjustment ($W = 19.0$, $p_{{holm}} = 0.054$, $r = 0.84$). Overall, these results indicate that AI first pass outputs achieve perceptual ratings comparable to human-authored baselines across all evaluated dimensions.


\begin{figure}[t]
 \centering
 \includegraphics[width=0.80\columnwidth]{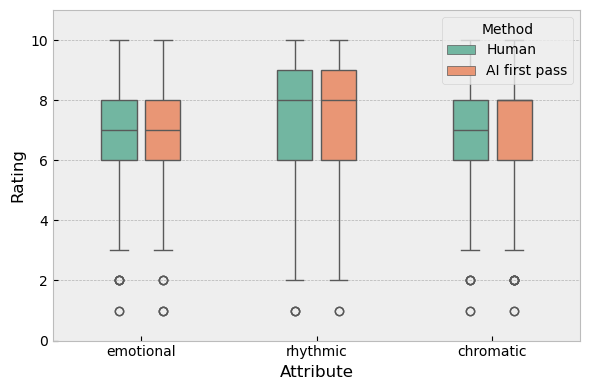}
 \caption{
  Box plots of individual participant ratings for AI-generated and human-generated outputs across 15 contexts and 32 participants for each perceptual attribute (emotional congruence, rhythmic synchronization, and chromatic congruence). Each box summarizes the distribution of ratings across all individual observations, illustrating the substantial overlap between AI and human conditions for all three attributes.
}
\vspace*{-5mm}
\label{fig:ratings-boxplot-all}
\end{figure}

\begin{table}[!h]
  \caption{
    Results of Wilcoxon signed-rank tests at individual rater level.
  }
  \label{tab:wilcoxon_stats_raterlevel}
  \scriptsize%
  \centering%
  \begin{tabu}{r * {4}{c}}
    \toprule
    Attribute & $W$ & $p$ & $p_{holm}$ & $r$ \\
    \midrule
    \texttt{emotional} & 31048 & 0.1488 & 0.2975 & 0.0659 \\
    \texttt{rhythmic}  & 29373 & 0.2099 & 0.2975 & 0.0572 \\
    \texttt{chromatic} & 25926 & 0.0013 & 0.0039 & 0.1468 \\
    \bottomrule
  \end{tabu}
\end{table}
\vspace*{-3mm}

\begin{figure}[!ht]
 \includegraphics[width=\linewidth]{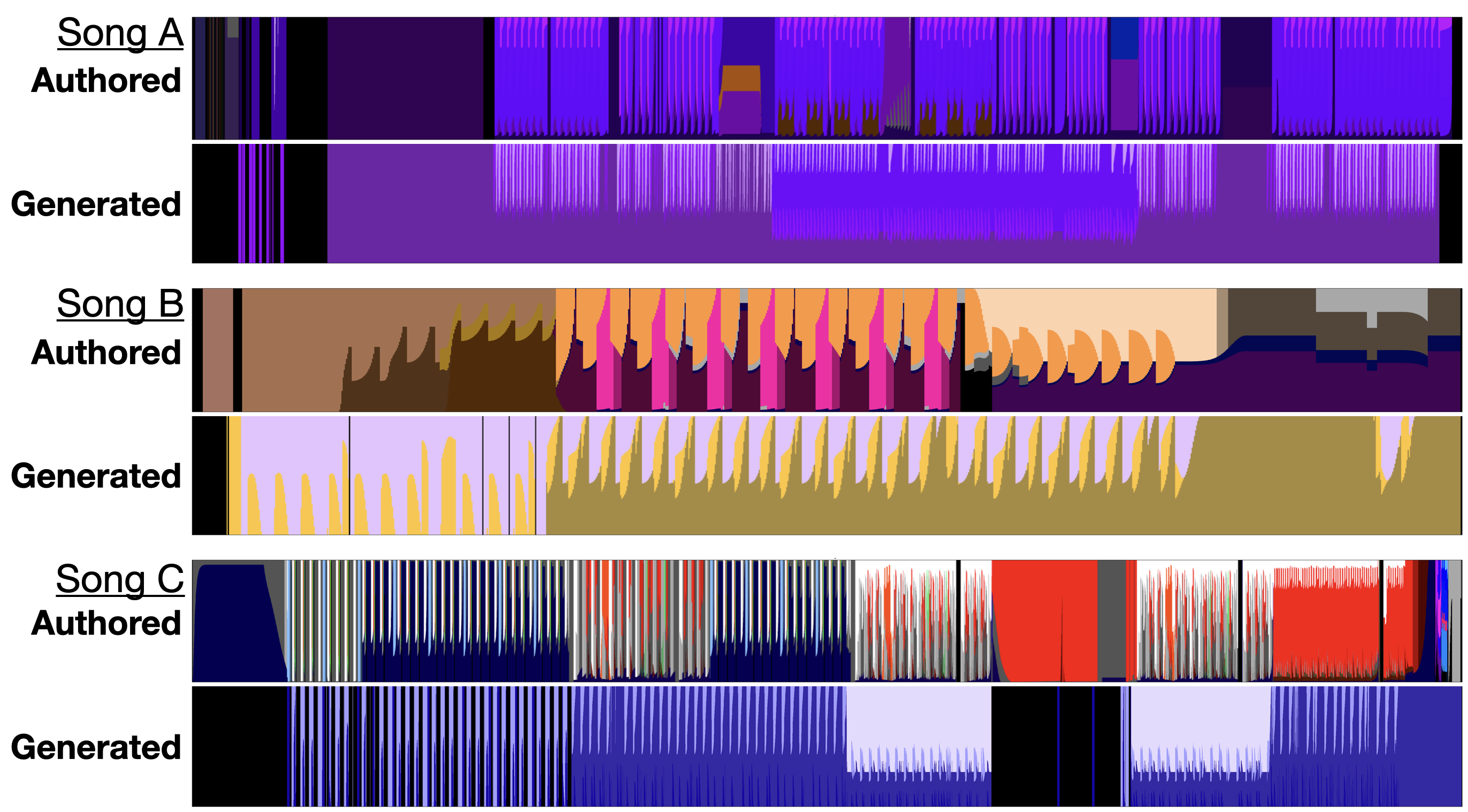}
 \caption{Visualizing the light designs of three songs, with each showing hand-authored and generated versions. The x axis shows frames along the songs' timeline, and the y axis shows the ratios of colors present during a given frame. Note how generated versions generally reflect section-structure, while having reduced color palette.}
 \label{fig:song_visualizations}
 \vspace*{-5mm}
\end{figure}

\cref{tab:wilcoxon_stats_raterlevel} presents an optional sensitivity analysis conducted at the individual rater level, in which Wilcoxon signed-rank tests were applied directly to paired AI first pass and human-authored ratings without aggregation across contexts. This analysis is reported for completeness and to assess the robustness of the primary findings, while acknowledging that individual ratings are not statistically independent. Interestingly, this study reveals a statistically significant difference for chromatic congruence after Holm correction ($p_{holm} = 0.0039$), while emotional and rhythmic attributes remain non-significant. When considering these findings, it is helpful to visualize the light content side by side. Figure \ref{fig:song_visualizations} shows the dominant colors for each frame along the songs' timelines. While this perspective hides details of moment-by-moment light activity, as well as how lighting is distributed spatially, it allows a high-level comparison of hand-authored and generated versions. This illustrates how (i) the start and end points of musical sections are largely consistent across both versions, (ii) the density of light activity (i.e. the number of light events within a section) are largely consistent across both versions, and (iii) general color palettes are adhered to in both versions. However, generated color palettes are generally less rich and diverse compared to hand-authored ones (e.g. in songs B and C).

Overall, the similarity of the visualizations illustrates how our system creates a solid first draft that matches hand-authored references. The system is able to adequately reflect the emotional content of the music by closely matching the rhythm and the color pallet. A human author, in the meanwhile, achieves the same emotional match without being overly explicit with color. The liberties taken in coloring are illustrated well in Figure \ref{fig:song_visualizations} where the manual versions of songs B and C feature additional colors that may not be as present in the video footage, but provide interesting variety.

It is important to note that this does not indicate participant \textit{preference} of one condition over the other, as there are aspects of the human-designed lightscapes that are not captured by the questions we asked. For example, some participants left qualitative comments describing the AI designs as ``repetitive", ``predictable", and ``too explicit" (regarding the translation of audio events to light cues). These impressions could still come with high ratings across the questions in Table \ref{tab:questions}. While these findings show that the system creates a viable baseline, we do not consider the two designs interchangeable.

%% file: 90_conclusion.tex
\section{Conclusion}

To address the skill and time constraints limiting designed light to professional contexts, this paper presented an system for the AI co-creation of ambient lightscapes for music videos. By modeling professional design heuristics, the system employs multi-modal analysis and synthesizes editable object-based light designs. Evaluation against hand-authored content confirms that the autonomous output serves as a viable baseline for further refinement by a designer, successfully reproducing the rhythmic, structural, and chromatic synchronization found in human-authored designs. This workflow demonstrates the potential for AI to facilitate the broader adoption of immersive, synchronized lighting experiences beyond professional venues.
Future work will examine how iterative human–AI refinement shapes creative outcomes and designer experience, and will compare co-created lightscapes against purely human-authored designs to assess the strengths and trade-offs of AI-assisted workflows.
